\documentclass[conference]{IEEEtran}
\IEEEoverridecommandlockouts

\usepackage{cite}
\usepackage{amsmath,amssymb,amsfonts}
\usepackage{algorithmic}
\usepackage{graphicx}
\usepackage{textcomp}
\usepackage{xcolor}
\usepackage{tabularx}
\usepackage{booktabs}
\usepackage{pifont}
\usepackage{multirow}
\usepackage{array}
\usepackage{hyperref}
\usepackage{float}
\usepackage[switch]{lineno}

\newcolumntype{Y}{>{\centering\arraybackslash}X}

\def\BibTeX{{\rm B\kern-.05em{\sc i\kern-.025em b}\kern-.08em
    T\kern-.1667em\lower.7ex\hbox{E}\kern-.125emX}}

\begin{document}


\title{AgenticVM: Agentic AI for Adaptive Software Vulnerability Management}

\author{%
    \begin{minipage}{1.0\textwidth}
        \centering
        Asrul Arifin$^{\ast}$,
        Hussain Ahmad$^{\ast\dagger}$\thanks{$\dagger$ Corresponding author},
        Yiyao Zhang$^{\ast}$,
        Diksha Goel$^{\ddagger}$  \\
        \smallskip
        \smallskip
        $^{\ast}$Adelaide University, Australia;
        $^{\ddagger}$CSIRO's Data61, Australia \\

        \smallskip
        \smallskip
        
        \{asrulsani.arifin, yiyao.zhang\}@student.adelaide.edu.au;
        hussain.ahmad@adelaide.edu.au;
        diksha.goel@csiro.au

    \end{minipage}
}

\maketitle

\begin{abstract}

As software systems grow in scale and complexity, vulnerability management is increasingly strained by high alert volumes, fragmented toolchains, and manual triage processes. We introduce AgenticVM, a multi-agent framework that integrates large language models with security tools to automate vulnerability detection, assessment, prioritization, and reporting. AgenticVM combines rule-based processing, a BERT-based CVSS prediction module, and specialised LLM-driven agents, leveraging data from sources such as the National Vulnerability Database and the European Union Vulnerability Database. Across multiple evaluation scenarios, AgenticVM reduces raw scanner outputs into compact, actionable queues, achieving up to 98\% alert reduction (e.g., from 3,983 findings to 82 high-priority items), while predicting missing CVSS attributes with 89.3\% accuracy. These results demonstrate improved prioritisation efficiency and reduced analyst workload without compromising risk visibility. Beyond performance, the framework provides practical design insights into agent decomposition, tool–LLM integration, and human-in-the-loop governance for real-world deployment.

\end{abstract}

\begin{IEEEkeywords}
Agentic AI, Cybersecurity, Vulnerability Management, LLMs, Natural Language Processing
\end{IEEEkeywords}

\section{Introduction}

The rapid expansion and increasing complexity of modern software systems have introduced a new wave of cybersecurity challenges \cite{jayalath2024microservice}, placing software vulnerabilities at the forefront of system risk and resilience \cite{ahmad2023review}. As organisations seek to develop secure and sustainable software, effective vulnerability management has become a critical priority \cite{abdulsatar2025towards}. Notable incidents such as the WannaCry ransomware attack \cite{b1} and the Heartbleed vulnerability disclosure \cite{b2} highlight how inadequately addressed vulnerabilities can escalate into large-scale disruptions with far-reaching consequences. In contemporary security operations centres (SOCs), vulnerability discovery consistently outpaces remediation efforts \cite{ullah2026skills}, leading to persistent backlogs that significantly elevate organisational risk and drive up operational costs.

Vulnerability management is therefore no longer a background technical task; it is a front-line operational function affecting service reliability, compliance, and business continuity. Yet in practice, vulnerability handling remains dominated by fragmented scanners, manually assembled reports, and human triage queues \cite{chopra2026chatnvd}. This challenge is amplified by the scale of modern vulnerability disclosure. In 2025 alone, the National Vulnerability Database (NVD) reported over 40,000 new CVEs \cite{b3}. This continued growth contributes to sustained remediation backlog, delayed patching, and prolonged exposure to exploitation. As a result, security teams are increasingly forced to prioritise under uncertainty, often relying on incomplete context and static severity scores.

Traditional software vulnerability management has relied on manual processes, rule-based systems \cite{b4}, and conventional machine-learning approaches \cite{b5}. Although these methods have improved parts of the analysis pipeline, they remain limited in several ways. Rule-based techniques are inherently static and brittle when facing evolving attacker behaviour or previously unseen report formats. Classical machine learning requires carefully curated features and extensive retraining, creating non-trivial maintenance and computational overhead. More recently, large language models (LLMs) and agentic AI systems have shown promise in complex reasoning and workflow automation \cite{chopra2026chatnvd}, but many current solutions still optimize for isolated subtasks rather than full operational integration. Agentic AI provides a practical path forward by enabling coordinated workflows that reduce noise and accelerate response while preserving practitioner control \cite{zhang2026explainable}. Rather than using a monolithic model for all decisions, agentic architectures decompose end-to-end vulnerability management into specialized components, each responsible for a distinct and auditable task. This enables stronger observability, better fault isolation, and more reliable governance.

This paper introduces \textbf{\textit{AgenticVM, an agentic AI framework designed for real-world vulnerability management}}. AgenticVM employs a modular, multi-agent architecture to support the entire lifecycle, from detection and assessment to severity prediction, prioritisation, and remediation recommendation. The framework integrates authoritative data sources such as the National Vulnerability Database (NVD) \cite{b3} and the European Union Vulnerability Database (EUVD) \cite{b7}, enabling it to automatically estimate missing CVSS scores, generate context-aware mitigation guidance, and facilitate human-in-the-loop oversight, observability, and post-deployment governance. Positioned as both a research contribution and an engineering design report, this work emphasises not only predictive performance but also architectural decisions, system decomposition, and practical deployment trade-offs relevant to real-world operational settings.

\begin{table*}[t]
\scriptsize
\centering
\caption{Comparison of AgenticVM with existing vulnerability management approaches}
\begin{tabularx}{\textwidth}{lccccccc}
\hline
\textbf{Study} &
\textbf{Detection} &
\textbf{Assessment} &
\textbf{Prioritization} &
\textbf{Recommendation} &
\textbf{CVSS Prediction} &
\textbf{Agentic AI} \\
\hline
Wu et al. (2025) \cite{b9}              & -- & \ding{51} & -- & -- & -- & \ding{51} \\
Narajala and Narayan (2025) \cite{b6}   & \ding{51} & -- & -- & \ding{51}  & -- & \ding{51} \\
Wali and Bulla (2024) \cite{b10}        & \ding{51} & -- & -- & -- & -- & \ding{51} \\
Toprani and Madisetti (2025) \cite{b11} & \ding{51} & -- & -- & \ding{51}  & -- & \ding{51} \\
R et al. (2025) \cite{b12}              & \ding{51} & -- & -- & \ding{51}  & -- & \ding{51} \\
Suggu (2025) \cite{b13}                 & \ding{51} & -- & -- & \ding{51}  & -- & \ding{51} \\
Li, Z et al. (2018) \cite{b14}          & \ding{51} & -- & -- & -- & -- & -- \\
Almotiri (2025) \cite{b15}              & \ding{51} & -- & -- & -- & -- & -- \\
Munoz et al. (2024) \cite{b16}          & \ding{51} & -- & -- & -- & -- & -- \\
Wartschinski et al. (2022) \cite{b17}   & \ding{51} & -- & -- & -- & -- & -- \\
Li, X et al. (2021) \cite{b18}          & \ding{51} & -- & -- & -- & -- & -- \\
Khazaei, Ghasemzadeh and Derhami (2016) \cite{b19} & -- & -- & -- & -- & \ding{51}  & -- \\
Aota et al. (2020) \cite{b20}           & -- & \ding{51} & \ding{51} & -- & -- & -- \\
Bozorgi et al. (2010) \cite{b20}        & -- & -- & \ding{51} & -- & -- & -- \\
Elbaz, Rilling and Morin (2020) \cite{b22} & -- & -- & -- & -- & \ding{51}  & -- \\
Gawron et al. (2018) \cite{b5}          & -- & -- & \ding{51} & -- & -- & -- \\
Han et al. (2017) \cite{b23}            & -- & -- & -- & -- & \ding{51}  & -- \\
Jacobs et al. (2020) \cite{b24}         & -- & -- & \ding{51} & -- & \ding{51}  & -- \\
Jiang et al. (2020) \cite{b25}          & -- & -- & -- & -- & \ding{51}  & -- \\
Le and Babar (2022) \cite{b26}          & -- & \ding{51} & -- & -- & \ding{51}  & -- \\
Ognawala et al. (2018) \cite{b27}       & -- & \ding{51} & -- & -- & -- & -- \\
Russo et al. (2019) \cite{b28}          & -- & -- & -- & \ding{51} & -- & -- \\
Sharma, Sibal and Sabharwal (2021) \cite{b29} & -- & -- & \ding{51} & -- & -- & -- \\
Zhang, H et al. (2024) \cite{b30}       & -- & \ding{51} & -- & -- & -- & -- \\
Zhang, L and Thing (2018) \cite{b31}    & -- & \ding{51} & \ding{51} & -- & -- & -- \\
Nie et al. (2022) \cite{b4}             & \ding{51} & -- & -- & -- & -- & -- \\
Jones and Omar (2023) \cite{b32}        & \ding{51} & -- & -- & -- & -- & -- \\
Chen et al. (2023) \cite{b33}           & \ding{51} & -- & -- & -- & -- & -- \\
Yang et al. (2022) \cite{b34}           & \ding{51} & -- & -- & -- & -- & -- \\
Tommy, Sundeep and Jose (2017) \cite{b35} & \ding{51} & -- & -- & -- & -- & -- \\
Gujar (2024) \cite{b36}                 & \ding{51} & -- & -- & -- & -- & -- \\
ZeMicheal et al. (2024) \cite{b37}      & -- & \ding{51} & -- & -- & -- & \ding{51} \\
Cepeda, Colome and Castrillon (2011) \cite{b38} & -- & \ding{51} & -- & -- & -- & -- \\
Gladkikh and Zakharov (2025) \cite{b39} & -- & -- & -- & -- & -- & \ding{51} \\
Aziz and Mohasseb (2024) \cite{b40}     & -- & \ding{51} & -- & -- & -- & -- \\
Russell et al. (2018) \cite{b41}        & \ding{51} & -- & -- & -- & -- & -- \\
Loevenich et al. (2025) \cite{b42}      & -- & \ding{51} & -- & -- & -- & -- \\
Liu et al. (2025) \cite{b43}            & -- & -- & -- & -- & -- & \ding{51} \\
Li, L et al. (2024) \cite{b44}          & \ding{51} & -- & -- & -- & -- & \ding{51} \\
Jie et al. (2025) \cite{b45}            & \ding{51} & -- & -- & -- & -- & \ding{51} \\
\textbf{This study (AgenticVM)}         & \ding{51} & \ding{51} & \ding{51} & \ding{51} & \ding{51} & \ding{51} \\
\hline
\end{tabularx}
\label{tab:comparison}
\end{table*}

Our paper makes the following contributions:\\

\begin{itemize}

    \item We propose AgenticVM, a modular multi-agent framework that integrates large language models with conventional security tooling to automate the full vulnerability management lifecycle, from detection and assessment to prioritisation and remediation.
    
    \item We develop a hybrid pipeline combining rule-based processing, a BERT-based CVSS prediction module, and specialised LLM-driven agents to enable reliable severity estimation, context-aware enrichment, and effective reduction of alert noise without compromising risk coverage.
    
    \item We demonstrate the effectiveness of AgenticVM through empirical evaluation, showing significant improvements in alert reduction and prioritisation efficiency, and distil key engineering insights on agent decomposition, tool–LLM integration boundaries, failure containment, observability, and human-in-the-loop governance for real-world deployment.
    
\end{itemize}

\section{Related Work}

This section reviews existing literature relevant to AgenticVM and establishes the motivation for its development. Although artificial intelligence has been extensively applied across domains \cite{abbas2025scalar}, such as software engineering \cite{ahmad2025towards, ahmad2024smart, ahmad2025resilient}, cybersecurity \cite{goel2025co, ahmad2025survey, goel2024machine}, finance \cite{zhang2025regimefolio, chen20253s}, education \cite{ahmad2025future, haque2022think}, and other critical environments \cite{jois2026australian, abbas2024robust, santhosh2026comparative}, its adoption in vulnerability management remains fragmented. Prior work typically addresses individual components, such as vulnerability detection, severity prediction, or prioritisation, in isolation, with limited focus on cohesive, end-to-end solutions. In the following, we provide a structured overview of these research directions and identify key gaps that underpin the need for a unified, agentic framework like AgenticVM.

\subsection{ML-Based Vulnerability Prioritisation}
Prior work has progressively shifted vulnerability management from static rule-based triage to automated pipelines. While scanner-driven approaches detect issues effectively, they still impose substantial manual prioritisation burdens on analysts. Several studies use machine learning to improve severity ranking, exploitability estimation, or risk classification from CVE-scale datasets. Among the reviewed works, Aota et al. \cite{b20}, Gawron et al. \cite{b5}, Jacobs et al. \cite{b24}, and Sharma et al. \cite{b29} investigate automated prioritisation or ranking mechanisms. More recent work, such as Abdulsatar et al. \cite{b59} explores advanced learning-based risk assessment for microservice environments. However, most such approaches optimise predictive accuracy in isolation and offer limited support for end-to-end workflow orchestration, observability, or human-in-the-loop decision-making.

\subsection{Deep Learning and NLP for Vulnerability Prediction}
Several studies apply deep learning and natural language processing to vulnerability analysis and CVSS prediction. Wartschinski et al. \cite{b17}, Li et al. \cite{b18}, Chen et al. \cite{b33}, and Yang et al. \cite{b34} use deep architectures for vulnerability classification or extraction. Khazaei et al. \cite{b19}, Elbaz et al. \cite{b22}, Han et al. \cite{b23}, Jiang et al. \cite{b25}, and Le and Babar \cite{b26} explicitly address CVSS prediction or severity inference. These studies show that textual vulnerability descriptions contain enough semantic signal for model-driven estimation of security impact. Nevertheless, many prior approaches treat prediction as an isolated classification problem and do not explain how predictions should be integrated into operational decision pipelines when official CVSS scores are missing or delayed.

\subsection{LLM-Based Security Tools and Agentic Systems}
LLM-based security tools support vulnerability explanation, remediation suggestion, and workflow assistance, but they also introduce challenges, including prompt sensitivity, cost, reproducibility, and governance concerns \cite{b13,b46}. Multi-agent systems further decompose complex tasks through coordinated interactions \cite{b57}. In cybersecurity, emerging studies such as Wu et al. \cite{b9}, Narajala and Narayan \cite{b6}, Toprani and Madisetti \cite{b11}, and Jie et al. \cite{b45} indicate that agentic architectures can support vulnerability assessment, orchestration, or remediation planning. However, evaluations remain largely task-specific and often lack full-lifecycle integration, explicit safety mechanisms, and operational governance.

\subsection{Research Gap and Positioning}
The key gap is not the absence of capable models, but the lack of reliable, end-to-end frameworks that integrate heterogeneous components, rules, classical ML, transformer models, and LLM-based agents, under real-world operational constraints. Existing approaches rarely support full-lifecycle coordination or provide strong mechanisms for observability, failure handling, and human oversight. AgenticVM addresses this gap through a multi-agent architecture that unifies detection, assessment, prioritisation, prediction, and recommendation within a controlled, deployable workflow for security operations. To contextualise this gap, Table~\ref{tab:comparison} compares representative studies across six core aspects of software vulnerability management.

\section{Research Methodology}

\begin{figure*}[t]
\centering
\includegraphics[width=0.62\textwidth]{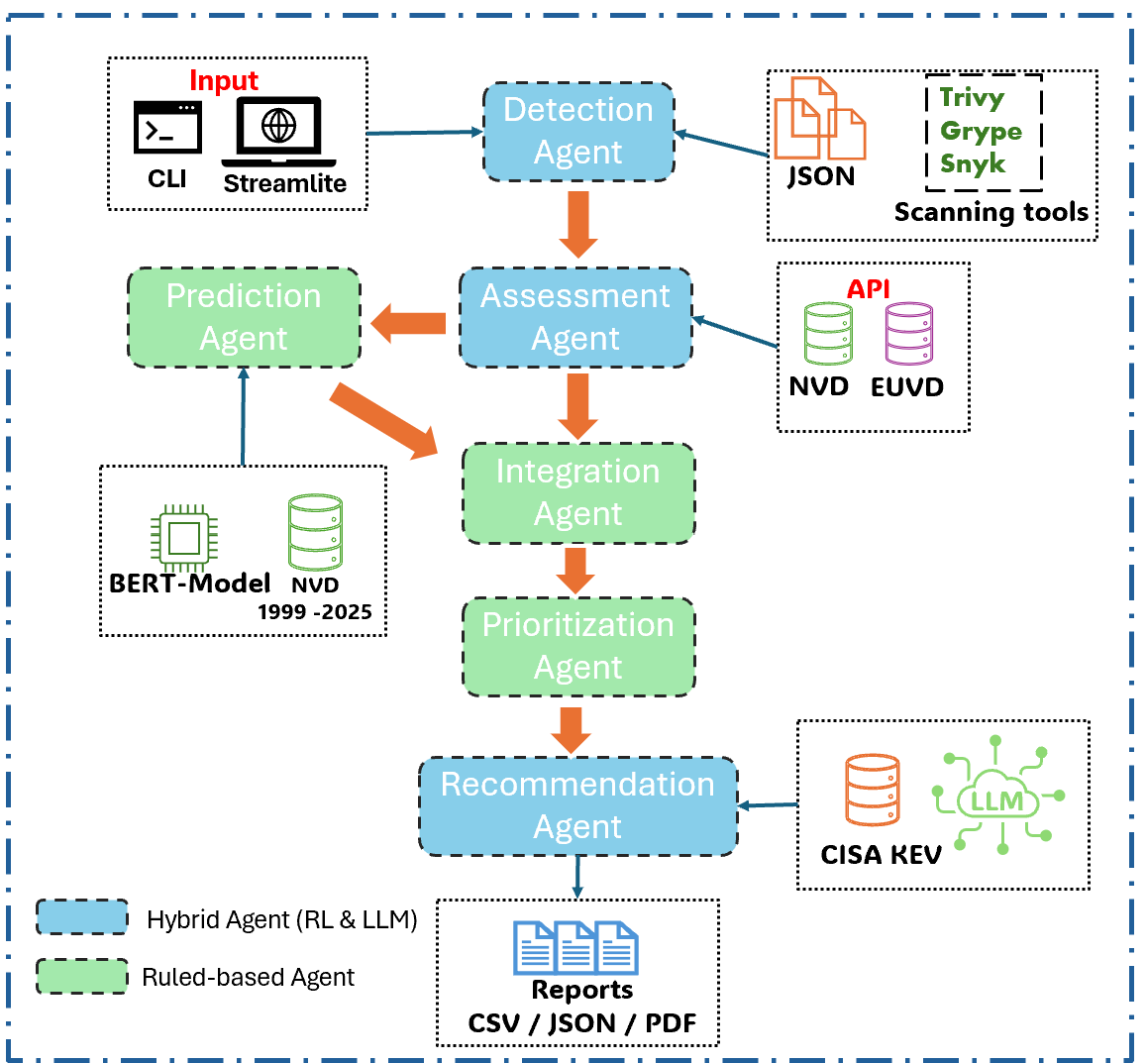}
\caption{AgenticVM Architecture illustrating the orchestration layer and the flow of vulnerability data across specialised agents.}
\label{fig:architecture}
\end{figure*}

\begin{figure*}[t]
\centering
\includegraphics[width=0.78\textwidth]{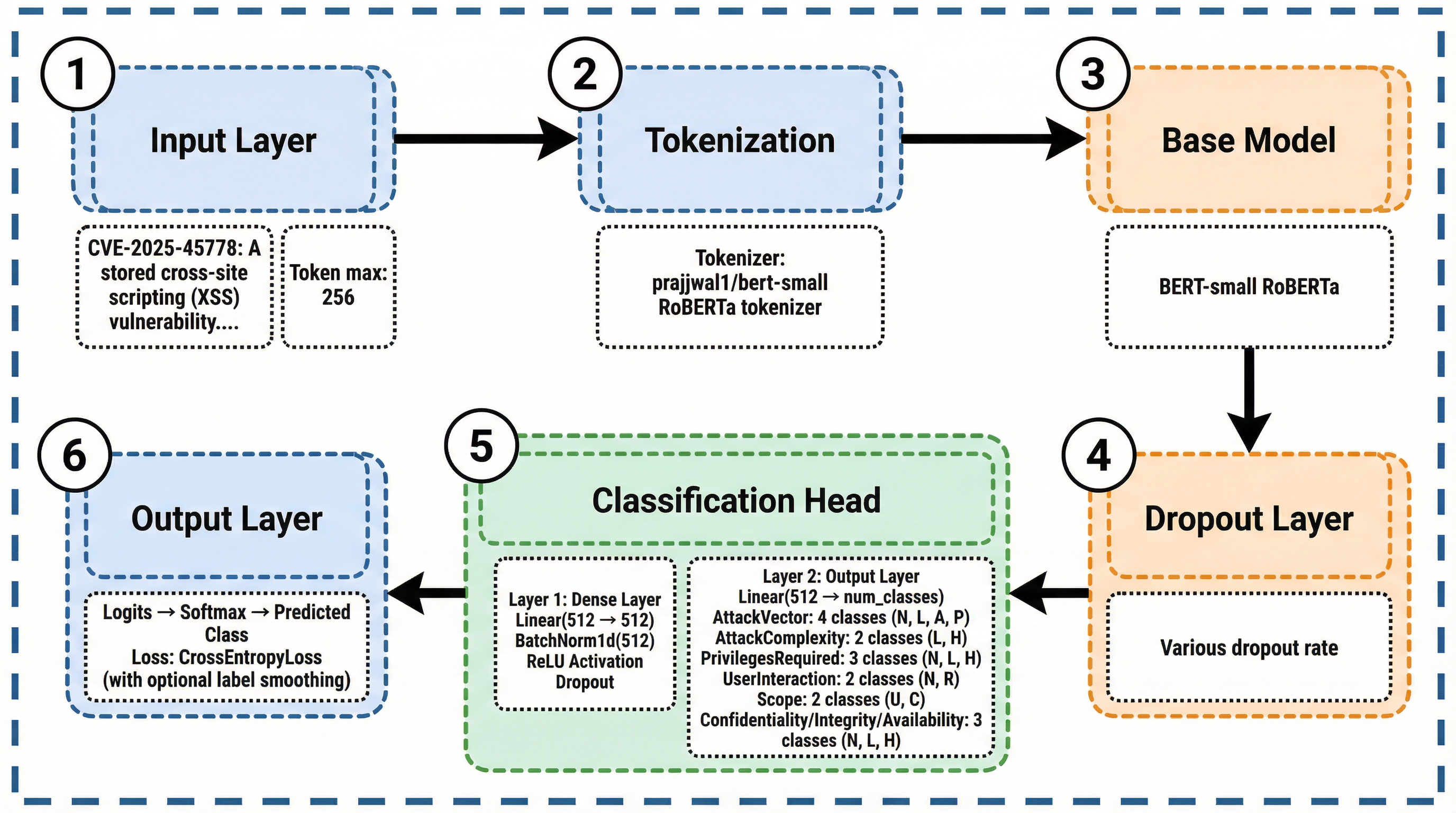}
\caption{BERT-based CVSS prediction model architecture.}
\label{fig:model-architecture}
\end{figure*}

This section outlines the methodology employed in developing AgenticVM. The methodology adopts a systematic approach that enables a thorough understanding of the problem space, the construction of a coherent system architecture, and the implementation of both the prediction model and the agentic components. It also provides a structured foundation for evaluating the performance and effectiveness of the proposed framework. The methodology is organised into five key phases: conducting the literature review, acquiring and preparing the data, designing the system architecture, implementing the framework, and finally carrying out the evaluation, analysis, and discussion.

\subsection{Phase 1: Literature Review}
The research begins with a comprehensive literature review to establish a strong theoretical and empirical foundation. This phase involves critically examining prior work in six primary domains: vulnerability detection, assessment, prioritisation, mitigation recommendation, CVSS prediction, and agentic AI. Special attention is given to studies that apply transformer-based models such as BERT and RoBERTa, as well as recent work on LLM-based cybersecurity workflows. Through this review, current limitations in software vulnerability management and gaps in existing CVSS prediction and orchestration approaches are identified. The insights gained from the literature directly inform the conceptualisation of AgenticVM and justify the research direction.

\subsection{Phase 2: Data Acquisition and Preparation}
The second phase focuses on collecting and preparing data required for both model development and system evaluation. Vulnerability descriptions and their corresponding CVSS scores are sourced primarily from the NVD \cite{b3} and EUVD \cite{b7}, along with supplementary datasets such as the CISA Known Exploited Vulnerabilities (KEV) catalog \cite{b48}. Raw data often contains inconsistencies, missing information, or duplicated records, making preprocessing essential. This phase includes data cleaning, text normalisation, label verification, and tokenisation. Additionally, the dataset is partitioned into training, validation, and test subsets following standard machine-learning procedures. To reduce the risk of data leakage, deduplication is performed before splitting, using CVE identifier as the primary key and normalised description text as a secondary near-duplicate check.

\subsection{Phase 3: System Architecture Design}
Building on the prior phases, the third phase focuses on designing the architecture of AgenticVM. This includes defining the roles and responsibilities of each agent within the multi-agent system, such as the detection agent, assessment agent, prediction agent, integration agent, prioritisation agent, and recommendation agent. Fig.~\ref{fig:architecture} illustrates the interactions, communication protocols, and decision flow among these agents. In parallel, the architecture for the BERT-based CVSS prediction model is specified, including the choice of transformer backbone, fine-tuning strategy, and confidence-gating policy represented in Fig.~\ref{fig:model-architecture}. The design phase ensures conceptual clarity and provides a blueprint for the subsequent implementation.

\subsection{Phase 4: Implementation}
The fourth phase involves translating the system design into an operational prototype. Each agent is implemented using LangGraph \cite{b47} as the orchestration framework, while the BERT-based prediction model is fine-tuned on the prepared dataset. This phase requires developing modules for text preprocessing, tokenisation, model inference, schema validation, and multi-agent communication logic. Attention is also given to the integration layer that connects the NLP model with the agent workflow, enabling end-to-end operation from vulnerability ingestion through automatic CVSS prediction to final recommendation. In the current implementation, rule-based modules are used where deterministic behaviour is preferable, while LLM-driven agents are reserved for unstructured interpretation and remediation synthesis.

\subsection{Phase 5: Evaluation, Analysis, and Discussion}
The final phase evaluates the effectiveness of both the machine-learning model and the multi-agent framework. For the CVSS prediction component, multiple performance metrics such as accuracy, F1-score, precision, and recall are computed to assess predictive reliability. These results are compared against alternative transformer configurations to determine the degree of improvement offered by the proposed approach. The multi-agent system is evaluated through scenario-based testing that examines Task Completion Time (TCT), Cost Efficiency Rate (CER), Token Consumption Rate (TCR), and Cooperation Performance ($C_{\text{perf}}$). In addition, lightweight baseline comparisons and ablation-style analysis are performed to assess the operational contribution of different architectural components.

Beyond numerical evaluation, this phase also includes a reflective analysis of the findings. Strengths, limitations, unexpected outcomes, and the practical implications of deploying such a system in real-world security operations are discussed. This discussion not only contextualises the results but also highlights opportunities for future enhancements, such as dynamic patch validation, broader datasets, trust-calibrated human-in-the-loop governance, and more resource-efficient deployment strategies.

\section{System Architecture and Engineering Design}

This section presents the concrete architectural decomposition used in AgenticVM and explains why each specialised agent is implemented with a specific technique, namely rule-based logic, learned model, or LLM component.

The multi-agent orchestration layer manages the complex data flow and decision logic between agents. Unlike a purely linear pipeline, the orchestrator employs conditional routing based on intermediate outputs, such as sending low-confidence predictions to expert review or flagging incomplete records for validation. Each component is engineered to handle specific data modalities while maintaining system-wide coherence.

\subsection{Detection Agent [Hybrid]}
The Detection Agent is designed to identify security-relevant findings from heterogeneous sources with high recall. It employs a bifurcated processing strategy. For industry-standard scanner outputs such as Trivy, Snyk, and Grype, it uses deterministic rule-based parsers to serialise JSON or text outputs into a canonical internal format, preserving package names, versions, CVE identifiers, and associated metadata. For unstructured CLI logs or non-standard textual findings, it leverages an LLM-based parsing routine using constrained prompting and schema-aware extraction. This allows the system to interpret inputs that would be difficult to handle reliably using regex or static templates alone.

\subsection{Assessment Agent [Rule-based]}
The Assessment Agent functions as a logical verification gate. Its primary purpose is to reduce false positives and alert fatigue by confirming that identified vulnerabilities are relevant to the target environment. It cross-references affected packages or components against available project context and, where applicable, against local asset metadata. Deterministic validation is preferred here because operational trust depends on transparent and repeatable filtering logic. This agent therefore emphasises correctness, provenance, and explainability over generative flexibility.

\subsection{Prediction Agent [Learned ML]}
The Prediction Agent orchestrates a BERT-based transformer model to infer missing CVSS metrics from vulnerability descriptions. This component is operationally important because many vulnerability records lack complete scoring at the time analysts must make prioritisation decisions. The model processes normalised vulnerability text, tokenises inputs to a fixed maximum sequence length, and passes them through a fine-tuned transformer backbone. A lightweight classification head then predicts the eight base CVSS v3.1 metrics.

The prediction module also includes confidence-aware behaviour. Softmax output probabilities are used as a proxy for confidence; low-confidence cases can be flagged for manual review instead of being treated as final labels. This design supports practical deployment by reducing overreliance on uncertain model outputs while still enabling earlier triage.

\subsection{Integration Agent [Rule-based]}
The Integration Agent consolidates outputs from the Detection, Assessment, and Prediction Agents into a complete vulnerability record. It acts as the central data normalisation and conflict-resolution layer. In the current implementation, the agent enforces strict schema validation using Pydantic-style structured records, ensuring that fields such as identifier, severity, source provenance, and recommendation context are complete and type-consistent before downstream use.

\subsection{Prioritisation Agent [Rule-based]}
In the current implementation, prioritisation is performed through deterministic CVSS-based tier assignment rather than through an unconstrained learned scorer. Vulnerabilities are first filtered using a configurable CVSS threshold, set to 7.0 in the reported evaluation configuration, and then mapped into transparent priority bands using official CVSS v3.1 severity ranges. Additional signals such as CISA KEV or exploit-likelihood context can be attached as analyst-facing enrichment, but they are not encoded as an opaque multiplier in this manuscript version. This choice improves reproducibility and aligns the prioritisation logic with the implemented pipeline.

\subsection{Recommendation Agent [LLM-driven]}
The Recommendation Agent uses LLM-assisted generation to synthesise remediation advice. It combines vulnerability context, known mitigation information, and authoritative references such as NVD or CISA KEV when available. In the current implementation, the agent follows a retrieval-grounded generation strategy so that recommendations remain tied to source material. The generated output is intended to support analyst decision-making, not to autonomously apply changes. Human approval remains required for destructive or production-impacting actions.

\section{Experimental Evaluation}

The experimental setup for this study was designed to ensure reproducibility, controlled evaluation, and realistic validation of the AgenticVM framework. This section outlines the computing environment, datasets, model features, evaluation metrics, experimental scenarios, and tools used throughout the experiments. The goal of this setup is to establish a reliable foundation for assessing both the prediction models and the agentic multi-agent system under practical software vulnerability management workflows.

\subsection{Experimental Setup}

\paragraph{Experimental Environment}
All experiments were conducted in a Windows 11 environment on a standard workstation configuration used for development and controlled replay. Local parsing, validation, and prioritisation components executed on-device, while language-intensive tasks used an external LLM. The language tasks used OpenAI \texttt{gpt-4o-mini-2024-07-18}: detection/parsing ran at temperature 0.0 with a maximum of 8{,}000 tokens, assessment at temperature 0.1 with a maximum of 20{,}000 tokens, and recommendation generation at temperature 0.1 with a maximum of 3{,}000 tokens, matching the pinned configuration documented in the replication package \cite{b50}.

\paragraph{Dataset Description}
The primary dataset used for training and evaluating the prediction models consists of CVE records aggregated from the National Vulnerability Database (NVD) \cite{b3}, the European Union Vulnerability Database (EUVD) \cite{b7}, and the CISA Known Exploited Vulnerabilities (KEV) catalog \cite{b48}. The resulting dataset includes 169{,}883 unique vulnerability records aligned with CVSS v3.1. Each record contains structured and unstructured fields such as CVE-ID, vulnerability descriptions, CVSS base scores, severity levels, publication dates, modification dates, and metadata related to attack vectors and impacts.

The dataset was divided using an 80/10/10 split into training, validation, and test sets, as shown in Table~\ref{tab:dataset_split}. Deduplication was performed before splitting, using CVE identifiers as the primary key and normalised description text as a secondary check for near duplicates. Split assignment was therefore CVE-level rather than row-level, reducing leakage risk.

\begin{table}[htbp]
\centering
\caption{Dataset split}
\label{tab:dataset_split}
\begin{tabular}{lll}
\toprule
Dataset & Count & Percentage \\
\midrule
Training Set   & 135{,}905 & 80\% \\
Validation Set & 16{,}989  & 10\% \\
Test Set       & 16{,}989  & 10\% \\
\midrule
Total          & 169{,}883 & 100\% \\
\bottomrule
\end{tabular}
\end{table}

\paragraph{Evaluation Metrics}
The performance of the prediction models and the agentic framework was assessed using a combination of classification metrics and agent-level operational metrics. For the transformer-based prediction models, evaluation focused on accuracy, precision, recall, and F1-score.

\begin{equation}
\text{Accuracy} = \frac{TP + TN}{TP + TN + FP + FN}
\label{eq:accuracy}
\end{equation}

\begin{equation}
\text{Precision} = \frac{TP}{TP + FP}
\label{eq:precision}
\end{equation}

\begin{equation}
\text{Recall} = \frac{TP}{TP + FN}
\label{eq:recall}
\end{equation}

\begin{equation}
F1 = 2 \times \frac{\text{Precision} \times \text{Recall}}
{\text{Precision} + \text{Recall}}
\label{eq:f1}
\end{equation}

For the multi-agent workflow, additional system-level metrics were used:

\begin{equation}
\text{TCT} = t_{\text{end}} - t_{\text{start}}
\label{eq:tct}
\end{equation}

\begin{equation}
\text{CER} = \frac{\text{Total successful tasks}}{\text{Total Cost (USD)}}
\label{eq:cer}
\end{equation}

\begin{equation}
\text{TCR} = \frac{\text{Total Tokens Consumed}}{\text{Total successful tasks}}
\label{eq:tcr}
\end{equation}

\begin{equation}
C_{\text{perf}} =
\left(
\frac{N_{\mathrm{successful\ handoffs}}}
{N_{\mathrm{expected\ handoffs}}}
\right)\times 100\%
\label{eq:cperf}
\end{equation}

Lower TCT and TCR are preferred because they indicate faster and more token-efficient execution, whereas higher CER and $C_{\text{perf}}$ indicate better cost efficiency and stronger inter-agent coordination.

\paragraph{Experimental Scenarios}
To evaluate the system's real-world applicability, experiments were performed across multiple software systems using vulnerability reports generated by widely adopted security scanning tools---Trivy, Grype, and Snyk. The selected open-source applications represent complementary workloads: \textit{online-boutique} \cite{b54}, \textit{train-ticket} \cite{b55}, and \textit{beer-shop} \cite{b56}. A controlled synthetic vulnerability report was also created to test prediction behaviour and validate integration with the CISA KEV catalog \cite{b48}.

In this study, a \textit{scenario} is defined as one end-to-end execution of the full workflow: detection $\rightarrow$ assessment $\rightarrow$ prediction $\rightarrow$ integration $\rightarrow$ prioritisation $\rightarrow$ recommendation. A scenario is considered successful when alerts are normalised, missing CVSS values are inferred or flagged, validated records are prioritised, and recommendations are generated with provenance.

\paragraph{Experiment Tools}
LangSmith was used for workflow tracing, providing execution logs and trace visualisation for debugging and inspection. A dedicated evaluation module was also developed to systematically compute all agentic performance metrics. This dual approach ensured that both low-level runtime traces and high-level quantitative performance summaries were captured consistently.

\subsection{Experimental Results}

\subsubsection{Performance of Classification Models}

The CVSS prediction component was evaluated across all eight CVSS base metrics using BERT-small and RoBERTa. The results from both models demonstrate strong predictive performance across multiple categories, with BERT-small showing slightly more stable overall behaviour.

\begin{table}[htbp]
\caption{BERT-small results}
\label{tab:smallbert}
\resizebox{\columnwidth}{!}{
\begin{tabular}{|l|llll|}
\hline
\multicolumn{1}{|c|}{\multirow{2}{*}{Vectors}} & \multicolumn{4}{c|}{BERT-small Model (\%)} \\ \cline{2-5}
\multicolumn{1}{|c|}{} & \multicolumn{1}{l|}{Accuracy} & \multicolumn{1}{l|}{Precision} & \multicolumn{1}{l|}{Recall} & F1-Score \\ \hline
Attack Vector (AV)         & \multicolumn{1}{l|}{93}   & \multicolumn{1}{l|}{93}   & \multicolumn{1}{l|}{93}   & 93 \\ \hline
Attack Complexity (AC)     & \multicolumn{1}{l|}{94}   & \multicolumn{1}{l|}{93}   & \multicolumn{1}{l|}{94}   & 93 \\ \hline
Privileges Required (PR)   & \multicolumn{1}{l|}{83}   & \multicolumn{1}{l|}{83}   & \multicolumn{1}{l|}{83}   & 83 \\ \hline
User Interaction (UI)      & \multicolumn{1}{l|}{93}   & \multicolumn{1}{l|}{93}   & \multicolumn{1}{l|}{93}   & 93 \\ \hline
Scope (S)                  & \multicolumn{1}{l|}{93}   & \multicolumn{1}{l|}{93}   & \multicolumn{1}{l|}{93}   & 93 \\ \hline
Confidentiality Impact (C) & \multicolumn{1}{l|}{86}   & \multicolumn{1}{l|}{86}   & \multicolumn{1}{l|}{86}   & 86 \\ \hline
Integrity Impact (I)       & \multicolumn{1}{l|}{86}   & \multicolumn{1}{l|}{86}   & \multicolumn{1}{l|}{86}   & 86 \\ \hline
Availability Impact (A)    & \multicolumn{1}{l|}{87}   & \multicolumn{1}{l|}{87}   & \multicolumn{1}{l|}{87}   & 86 \\ \hline
Overall                    & \multicolumn{1}{l|}{89.3} & \multicolumn{1}{l|}{89.3} & \multicolumn{1}{l|}{89.4} & 89.1 \\ \hline
\end{tabular}
}
\end{table}

\begin{table}[htbp]
\caption{RoBERTa results}
\label{tab:roberta}
\resizebox{\columnwidth}{!}{
\begin{tabular}{|l|llll|}
\hline
\multicolumn{1}{|c|}{\multirow{2}{*}{Vectors}} & \multicolumn{4}{c|}{RoBERTa Model (\%)} \\ \cline{2-5}
\multicolumn{1}{|c|}{} & \multicolumn{1}{l|}{Accuracy} & \multicolumn{1}{l|}{Precision} & \multicolumn{1}{l|}{Recall} & F1-Score \\ \hline
Attack Vector (AV)         & \multicolumn{1}{l|}{93}   & \multicolumn{1}{l|}{93}   & \multicolumn{1}{l|}{93}   & 93 \\ \hline
Attack Complexity (AC)     & \multicolumn{1}{l|}{91}   & \multicolumn{1}{l|}{96}   & \multicolumn{1}{l|}{93}   & 95 \\ \hline
Privileges Required (PR)   & \multicolumn{1}{l|}{83}   & \multicolumn{1}{l|}{83}   & \multicolumn{1}{l|}{83}   & 83 \\ \hline
User Interaction (UI)      & \multicolumn{1}{l|}{93}   & \multicolumn{1}{l|}{93}   & \multicolumn{1}{l|}{93}   & 93 \\ \hline
Scope (S)                  & \multicolumn{1}{l|}{91}   & \multicolumn{1}{l|}{92}   & \multicolumn{1}{l|}{91}   & 92 \\ \hline
Confidentiality Impact (C) & \multicolumn{1}{l|}{86}   & \multicolumn{1}{l|}{86}   & \multicolumn{1}{l|}{86}   & 86 \\ \hline
Integrity Impact (I)       & \multicolumn{1}{l|}{86}   & \multicolumn{1}{l|}{86}   & \multicolumn{1}{l|}{86}   & 86 \\ \hline
Availability Impact (A)    & \multicolumn{1}{l|}{86}   & \multicolumn{1}{l|}{86}   & \multicolumn{1}{l|}{86}   & 86 \\ \hline
Overall                    & \multicolumn{1}{l|}{88.6} & \multicolumn{1}{l|}{89.4} & \multicolumn{1}{l|}{88.9} & 89.3 \\ \hline
\end{tabular}
}
\end{table}

BERT-small achieved strong predictive capability, with accuracies ranging from 83\% to 94\% across individual metrics and an overall accuracy of 89.3\%. The strongest results were observed for Attack Vector, Attack Complexity, User Interaction, and Scope. RoBERTa exhibited a similar pattern, achieving an overall accuracy of 88.6\%, but with especially high precision in Attack Complexity. Based on this result, BERT-small was selected as the primary prediction model because it provided slightly better overall stability for the dataset used in this study.

\subsubsection{Scenario-Based Processing Outcomes}

\begin{table}[htbp]
\centering
\caption{AgenticVM scenario results (CVEs number)}
\label{tab:vumantic_scenario}
\resizebox{\columnwidth}{!}{
\begin{tabular}{@{}l p{1.5cm} llll p{1.5cm}@{}}
\toprule
\multicolumn{2}{l}{Stage}
& \multicolumn{1}{p{1.5cm}}{\centering Online boutique}
& Train ticket
& Beer shop
& \multicolumn{1}{p{1.5cm}}{\centering Prediction test} \\
\midrule
\multicolumn{2}{l}{1. Detection:}      &       &      &      &      \\
& Raw detection                        & 118   & 3,983 & 1,746 & 6 \\
& Unique CVEs                          & 34    & 155   & 146   & 6 \\
\multicolumn{2}{l}{2. Assessment:}    &       &       &       &   \\
& NVD                                  & 34    & 155   & 146   & 4 \\
& EUVD                                 & 34    & 144   & 135   & 4 \\
& Needs prediction                     & 0     & 0     & 0     & 2 \\
\multicolumn{2}{l}{3. Prediction:}    &       &       &       &   \\
& Predicted                            & 0     & 0     & 0     & 2 \\
& Failed                               & 0     & 0     & 0     & 0 \\
\multicolumn{2}{l}{4. Integration:}   &       &       &       &   \\
& Total integrated                     & 34    & 155   & 146   & 6 \\
& With CVSS                            & 34    & 155   & 146   & 6 \\
\multicolumn{2}{l}{5. Prioritization:}&       &       &       &   \\
& Prioritized                          & 17    & 82    & 82    & 4 \\
& Below Threshold                      & 17    & 73    & 64    & 2 \\
\multicolumn{2}{l}{6. Recommendation:}&       &       &       &   \\
& From CISA                            & 0     & 4     & 1     & 2 \\
& From LLM                             & 17    & 78    & 81    & 2 \\
& Total                                & 17    & 82    & 82    & 4 \\
\bottomrule
\end{tabular}
}
\end{table}

The scenario-based evaluation provides insight into how AgenticVM behaves in practical operational environments. Across the Online Boutique, Train Ticket, Beer Shop, and Prediction Test scenarios, the system successfully processed large volumes of raw detections and reduced them to unique CVE entries through deduplication and consolidation. For example, the Train Ticket scenario began with 3,983 detections but was reduced to 155 unique CVEs, demonstrating the system's ability to eliminate noise and collapse repeated scanner findings into a manageable vulnerability set.

Assessment results show that most vulnerabilities could be mapped directly using authoritative databases such as NVD and EUVD, with only the Prediction Test scenario requiring model-based completion for two CVEs. The integration stage confirmed that all validated or predicted vulnerabilities were successfully incorporated into the final dataset. Prioritisation then filtered high-severity vulnerabilities, reducing the final actionable set by approximately half in most scenarios. Finally, the recommendation stage demonstrated that AgenticVM can provide actionable guidance using either external authoritative references or LLM-generated remediation advice.

\subsubsection{Baseline Comparison}

\begin{table*}[htbp]
\centering
\caption{Lightweight baseline comparison for Online Boutique and Train-Ticket}
\label{tab:baseline_comparison}
\resizebox{\textwidth}{!}{%
\begin{tabular}{llcccc}
\toprule
\textbf{Scenario} & \textbf{Workflow} & \textbf{Raw Detections} & \textbf{Unique CVEs} & \textbf{Prioritized Items} & \textbf{Alert Reduction} \\
\midrule
\textbf{Online Boutique} & Baseline 1: Scanner$\rightarrow$Manual & 32 & 32 & 32 & 0\% \\
\textbf{Online Boutique} & Baseline 2: Single-Agent LLM & 32 & 32 & 28 & 12.5\% \\
\textbf{Online Boutique} & AgenticVM & 32 & 9 & 6 & 81.3\% \\
\midrule
\textbf{Train-Ticket} & Baseline 1: Scanner$\rightarrow$Manual & 3{,}983 & 3{,}983 & 3{,}983 & 0\% \\
\textbf{Train-Ticket} & Baseline 2: Single-Agent LLM & 3{,}983 & 3{,}210 & 2{,}847 & 28.5\% \\
\textbf{Train-Ticket} & AgenticVM & 3{,}983 & 155 & 82 & 97.9\% \\
\bottomrule
\end{tabular}%
}
\end{table*}

For baseline comparison, two lightweight reference workflows were used: a non-agentic scanner$\rightarrow$manual workflow, and a single-agent LLM workflow that combines parsing, scoring, and recommendation in one prompt-driven process. Using identical scenario artefacts, AgenticVM reduced raw detections to much smaller high-risk queues. In Online Boutique, the framework reduced 32 detections to 6 prioritised items, and in Train-Ticket it reduced 3,983 detections to 82 prioritised items. The scanner$\rightarrow$manual baseline retained full alert load, while the single-agent LLM workflow achieved only modest reduction and demonstrated weaker provenance control and higher prompt sensitivity. This comparison should be interpreted as a lightweight process-level benchmark rather than a full exhaustive performance benchmark against all possible alternatives.

\subsubsection{Ablation-Style Contribution Summary}

\begin{table*}[htbp]
\centering
\caption{Ablation-style contribution summary (Online Boutique)}
\label{tab:ablation_summary}
\begin{tabular}{lccccc}
\toprule
\textbf{Variant} & \textbf{Unique CVEs} & \textbf{Prioritized} & \textbf{Time (s)} & \textbf{Accuracy} & \textbf{F1-Score} \\
\midrule
\textbf{Full System (AgenticVM)}      & 9  & 6 & 139.8 & 0.2968 & 0.4613 \\
\textbf{Without Prediction Agent}     & 9  & 4 & 128.5 & 0.2541 & 0.3982 \\
\textbf{Without Assessment Agent}     & 14 & 9 & 151.5 & 0.2868 & 0.4457 \\
\textbf{Without Prioritization Agent} & 9  & 9 & 206.0 & 0.2868 & 0.4457 \\
\textbf{Without Recommendation Agent} & 9  & 6 & 172.2 & 0.2868 & 0.4457 \\
\bottomrule
\end{tabular}%
\end{table*}

The ablation-style analysis highlights the role of each architectural component. Removing the Prediction Agent reduces overall accuracy and F1-score because vulnerabilities with missing or partial CVSS attributes can no longer be inferred and are conservatively excluded from the high-priority queue, leading to under-prioritisation. Although latency decreases slightly due to skipping model inference, this comes at the cost of weaker triage coverage. Removing the Assessment Agent increases the number of unique and prioritised items because false-positive filtering is bypassed. Removing the Prioritisation Agent causes all validated CVEs to be retained in the final queue, which increases downstream effort and total execution time. Removing the Recommendation Agent preserves the prioritised set but reduces remediation usability and increases fallback handling overhead.

The ablation summary should be interpreted as a scenario-specific operational sensitivity analysis, not as evidence that every component must change the same headline counts in every workload. In the Online Boutique scenario, some agents primarily affect verification effort, queue ordering, or usability rather than final count alone.

\subsubsection{Repeated-Run Stability}
To assess runtime stability, five controlled reruns of the full AgenticVM pipeline were conducted on the Online Boutique scenario. The system demonstrated robust performance stability with minor variations typical of probabilistic LLM-assisted workflows: Unique CVEs ($8.8 \pm 0.4$), Prioritized Items ($5.8 \pm 0.4$), Task Completion Time ($144.6\,\mathrm{s} \pm 7.7\,\mathrm{s}$), Accuracy ($0.2892 \pm 0.0152$), and F1-Score ($0.4518 \pm 0.0189$). Here, Accuracy and F1 refer to workflow-level agreement against scenario ground-truth triage decisions rather than the eight-way CVSS vector prediction metric used for the BERT model. These values are therefore most meaningful when interpreted comparatively across variants rather than in isolation.

\section{Discussion}

The results support three complementary interpretations of AgenticVM: predictive utility, workflow compression, and operational feasibility.

\subsection{Predictive Utility of the CVSS Model}
The BERT-small prediction component delivered strong performance, achieving 89.3\% overall accuracy across the eight CVSS base metrics. This result indicates that textual vulnerability descriptions provide sufficient signal for reliable inference of missing severity attributes in many cases. From an operational perspective, this is important because analysts often need to prioritise vulnerabilities before authoritative scores are fully available. In AgenticVM, model predictions are not treated as unquestionable ground truth; rather, they serve as decision support signals, particularly when combined with confidence-aware review policies.

\subsection{Workflow Compression and Alert Reduction}
A major contribution of AgenticVM is its ability to compress large volumes of scanner output into substantially smaller and more actionable queues. This effect is most clearly seen in the Train-Ticket and Online Boutique scenarios. The framework reduced raw detections to unique CVEs through deduplication, validation, and structured integration, and then further reduced the queue through deterministic prioritisation. The lightweight baseline comparison shows that this compression substantially exceeds what is achieved by a manual queue or a single-agent LLM workflow under the same input artefacts.

From a practitioner perspective, this matters because time-to-remediation is constrained not only by patch availability but also by analyst attention. Reducing the number of alerts that require immediate review helps teams focus effort on vulnerabilities most likely to affect operational risk.

\subsection{Engineering Lessons}
Several engineering lessons emerge from the study.

First, decomposing vulnerability management into specialised agents is more maintainable than relying on one monolithic LLM because it improves fault isolation, debugging, and operational control. Deterministic components are appropriate where explainability and reproducibility are critical, such as schema validation, threshold-based prioritisation, and relevance filtering. LLM-based components are most useful where input variability and natural-language interpretation dominate, such as unstructured log parsing or remediation synthesis.

Second, tool--LLM boundaries matter. The framework performs better when external tools and authoritative data sources are used for facts, while LLMs are constrained to interpretation, transformation, or recommendation tasks. This design reduces hallucination risk and improves provenance.

Third, observability is necessary for operational trust. Trace logging, schema checks, and stage-level metrics make it easier to understand where failures occur and to distinguish model uncertainty from workflow failure. These properties are especially important in environments where security decisions must be auditable.

\subsection{Failure Modes and Safeguards}
Four safeguards anchor the design: schema-checked parsing with fallback review, confidence-gated CVSS prediction, grounded remediation with approval gates, and mandatory-field validation for incomplete records. Across the reported scenarios, these controls contained errors without blocking workflow progress and preserved accountability. The framework therefore supports a more controlled deployment profile than an unconstrained end-to-end LLM workflow.

\subsection{Operational Implications for Security Teams}
AgenticVM reduces vulnerability triage burden through automated ingestion, severity estimation, prioritisation, and recommendation. Its 89.3\% CVSS inference accuracy supports earlier risk assessment before official scores are published, while destructive actions still require strict human-in-the-loop approval to preserve accountability. The system is therefore best understood as an analyst-amplification framework rather than a fully autonomous remediation engine.

\section{Threats to Validity}

Despite the strengths of the proposed approach, several limitations should be acknowledged. First, the experiments rely on a combination of local processing and external LLM services. Although the workflow is controlled and reproducible within the provided configuration, operational results such as latency and token consumption may vary depending on provider behaviour, model versioning, or infrastructure load. This affects external validity for direct performance replication across environments. Second, the dataset used for training and evaluation may not fully represent the diversity of real-world vulnerability descriptions. Although the data has been curated and deduplicated, it may still contain linguistic regularities or reporting conventions that differ from those encountered in operational environments. As a result, prediction quality may decrease when the model is exposed to unseen or noisier input distributions. Third, some reported measures, particularly workflow-level accuracy and F1 in the ablation-style analysis, are scenario-level agreement metrics rather than conventional classifier metrics. These are meaningful for internal comparison but should not be conflated with the CVSS vector prediction results. Care is therefore required when interpreting numerical values across different experimental layers. Fourth, the baseline comparison is intentionally lightweight. It is designed to illustrate operational differences between a manual queue, a single-agent LLM workflow, and the proposed agentic architecture, but it does not exhaustively benchmark all competing systems or optimisation strategies. Additional comparative studies would strengthen construct validity.

\section{Conclusion and Future Work}

This paper presented an integrated agentic multi-agent framework designed to automate key stages of software vulnerability management, from data acquisition and CVE extraction to CVSS prediction, prioritisation, and vulnerability recommendation. By combining deterministic parsing and validation, LLM-based agents, and a lightweight transformer prediction model, the system demonstrates how automated reasoning can reduce manual workload while maintaining reliable operational performance. The findings indicate that the prediction component, powered by a BERT-small model, achieved strong overall accuracy, while the multi-agent workflow proved effective in coordinating tasks such as CVE identification, data structuring, risk scoring, and remediation support. Most importantly, the system reduced large scanner outputs into substantially smaller prioritised queues, supporting more focused analyst action. Collectively, these results highlight the feasibility of using agentic AI to support vulnerability analysis in a more scalable, consistent, and timely manner.

Future work should connect recommended fixes to ephemeral sandbox environments, regression tests, policy checks, and staged rollout gates before production approval, extending recent work on automated vulnerability detection, remediation workflows, and agent governance. A second priority is human-centred evaluation of trust calibration, explanation usefulness, override behaviour, and team coordination in HITL workflows, particularly in agentic security systems. Further research is also needed on cost-aware orchestration, hybrid local/remote inference strategies, and robustness across diverse report formats and software ecosystems.

\section{Replication Package}

We provide a replication package containing the full implementation of the AgenticVM framework, experiment scripts, and all datasets used in this study. The project source code and materials are available in our GitHub repository at \cite{b50}, enabling researchers and practitioners to explore, validate, or extend the system. The resources are shared to encourage reuse, foster comparative studies, and support further development in Agentic AI for vulnerability management.

\clearpage
\bibliographystyle{IEEEtran}
\bibliography{references}

\clearpage
\appendices

\vspace{12pt}

\end{document}